\documentclass[twocolumn]{openjournal} 

\usepackage{orcidlink} 
\usepackage[T1]{fontenc}
\usepackage{ae,aecompl}
\usepackage[T1]{fontenc}
\usepackage{ae,aecompl}
\usepackage{hyperref}
\usepackage{url}

\usepackage{longtable}
\usepackage{booktabs}
\usepackage{tabu}
\usepackage{graphicx}
\usepackage{multirow}
\usepackage{ulem}
\usepackage{color}
\usepackage{amsmath}

\def \cm{~\rm{cm}}
\def \s{~\rm{s}}
\def \km{~\rm{km}}

\def \K{~\rm{K}}

\def \g{~\rm{g}}

\def \AU{~\rm{AU}}
\def \erg{~\rm{erg}}

\def \yr{~\rm{yr}}

\usepackage{xcolor}
\definecolor{redak}{rgb}{0.9,0.15,0.05}

\shorttitle{The failed scenario of M31-2014-DS1i}
\shortauthors{Soker}

\graphicspath{{./}{figures/}}

\begin{document}

\title{The failed failed-supernova scenario of M31-2014-DS1}

\author{Noam Soker\,\orcidlink{0000-0003-0375-8987}} 
\affiliation{Department of Physics, Technion - Israel Institute of Technology, Haifa, 3200003, Israel; soker@technion.ac.il}


\begin{abstract}
I examine a recently proposed failed-supernova scenario for the fading of the yellow supergiant event M31-2014-DS1, and find that it requires unlikely fine-tuned parameters to work, if at all. In the failed-supernova scenario, most of the yellow supergiant collapsed to form a black hole. Due to the energy carried by neutrinos from the cooling, collapsing core, gravity decreases, leading to the ejection of a small fraction of the outer envelope, some of which remains bound. The fallback accreted gas possesses large angular-momentum fluctuations due to the pre-collapse envelope convection. The fallback material forms intermittent accretion disks around the black hole that launch jets (or disk wind), which unbind most of the bound material. The failed-supernova scenario for M31-2014-DS1 requires that only $<1\%$ of the bound material be accreted by the black hole, but the jets do not shut down the backflow for over 10 years. I find this fine-tuned requirement unlikely. I also find that, due to the rapid radiative cooling of the outflow interaction zone with the outer gas, the expected radiation is about an order of magnitude or more above the observed value. These, as well as earlier challenges raised against the failed-supernova scenario, make the alternative type II intermediate-luminosity optical transient scenario, in which fading is due to dust ejection in a violent binary interaction, more likely.  The fading event M31-2014-DS1 does not support the failed-supernova scenario predicted by the neutrino-driven explosion mechanism of core-collapse supernovae.   
\end{abstract}

\keywords{Supernovae  -- Massive stars	--  Circumstellar material -- binary systems}



\section{Introduction} 
\label{sec:intro}

The two intensively studied core-collapse supernova (CCSN) explosion mechanisms, the jittering jets explosion mechanism (JJEM; \citealt{Soker2024UnivReview, Soker2025Learning}, for recent reviews) and the neutrino-driven (delayed-neutrino) mechanism (e.g., \citealt{Burrowsetal2024, Janka2025}), differ on their prediction of failed CCSN and black hole formation. The prediction of the neutrino-driven mechanism is that a non-negligible fraction of massive stars do not explode; their core collapses to form a black hole in a `failed supernova', accompanied by only a faint transient event (e.g., \citealt{AntoniQuataert2023}). Not all black holes are born in failed-supernovae in the neutrino-driven mechanism; many are formed in supernovae (e.g., \citealt{BurrowsWangVartanyan2025}).   
In contrast, in the JJEM, there are no failed CCSNe.
   
\cite{Adamsetal2017} studied the fading of the event N6946-BH1 \citep{Gerkeetal2015}, and attributed it to a failed supernova in the framework of the neutrino-driven mechanism. \cite{Humphreys2019} suggested that it was a yellow hypergiant on a post-red supergiant track, likely experiencing high mass loss before the collapse. 
We \citep{KashiSoker2017, SokerTypeII2021, BearetalTypeII2022} proposed an alternative explanation in the framework of the JJEM:  a type II intermediate luminosity optical transient (ILOT) event, where the merger of two stars expels dusty equatorial ejecta that obscures the merger remnant, or the binary stellar system if the two stars did not merge. \cite{Beasoretal2024} claimed that the type II ILOT scenario better fits their detection of a luminous infrared source at the position of N6946-BH1 (for a view in support of the failed-supernova scenario for N6946-BH1 see \citealt{Kochanek2024, Kochaneketal2024}). Another challenge to the claim of a neutrino-driven mechanism for many failed supernovae is recent studies suggesting that there are, at best, only a small number of failed supernovae \citep{ByrneFraser2022, StrotjohannOfekGalYam2024, Beasoretal2025, Healyetal2025}.  In addition, \cite{Beasoretal2025} argued that the luminosity inferred for many CCSN progenitors is underestimated, implying that many CCSN progenitors have masses of $ > 20 M_\odot$, reducing the fraction of failed CCSN to be very low (zero according to the JJEM).

Another failed-supernova candidate is M31-2014-DS1 as proposed by  \cite{Deetal2024}, who also described a failed-supernova scenario for this event (see also \citealt{Antonietal2025}). 
In \cite{Soker2024UnivReview}, I questioned their conclusion for the following reasons. 
(1) They assumed a spherically symmetric dust distribution, whereas in type II ILOTS it is not \citep{KashiSoker2017, SokerTypeII2021}.  \cite{Beasoretal2025} further support the non-spherical dust distribution of M31-2014-DS1. 
(2) The progenitor model \cite{Deetal2024} discuss is of a star with an initial mass of $M_{\rm ZAMS} \simeq 20 M_\odot$. At explosion, the mass is only $M \simeq 6.7 M_\odot$, with a radius of $\simeq 400 R_\odot$. To lose two-thirds of its mass, it had a binary interaction, because a single star of this mass cannot lose such a large mass (e.g., \citealt{Beasoretal2020, Zapartasetal2025, Gilkisetal2025}). Such an interaction, I argued, spun up the progenitor to a degree that the outer layers of the progenitor would form an accretion disk around the black hole. Such a disk would launch energetic jets, resulting in a bright explosion. 
(3) I further argued that even if there was no rotation, the envelope convection is vigorous enough to form intermittent accretion disks around the black hole (e.g., \citealt{Quataertetal2019, AntoniQuataert2022, AntoniQuataert2023}).  \cite{Deetal2024} noticed that the $\simeq 0.15 M_\odot$ outer envelope has the properties to form intermittent accretion disks. In their model, the ejected mass in the event is $\simeq 0.1 M_\odot$. Even if only $0.01 M_\odot$ is accreted through intermittent accretion disks and launches jets, namely, the JJEM, and releases $5 \%$ of its mass, the explosion energy is $0.0005 M_\odot c^2 \simeq 10^{51} \erg$; There is an explosion in the frame of the JJEM even in the formation of a black hole  (e.g., \citealt{Gilkisetal2016}).    
Following these points, in \cite{Soker2024UnivReview} I predicted that M31-2014-DS1 would reappear in several years.

Two recent papers study the fading event of the yellow supergiant M31-2014-DS1. \cite{DeKetal2026} claim that this event was a failed supernova, while \cite{Beasoretal2026} claim it is more likely to be a type II ILOT. 
While \cite{DeKetal2026} considered only the failed-supernova scenario and ignored the type II ILOT alternative scenario, \cite{Beasoretal2026} presented a thorough discussion and considered the two scenarios. 
\cite{Beasoretal2026} presents a discussion of previous claims for failed supernovae, including the explanation that there are no red supergiant (RSG) stars missing from the observed sample of II-P progenitors (e.g., \citealt{Beasoretal2025}). 
In this study, I show that the fallback accretion that is expected to power the event for years in the failed-supernova scenario is inconsistent with observations.

\section{The flow properties}
\label{sec:Jets}

\cite{Beasoretal2026} present new JWST, Submillimeter Array (SMA), and Chandra observations of M31-2014-DS1. There are no detections from the SMA and Chandra, yielding only upper limits on the fluxes; \cite{DeKetal2026} also obtained an upper limit on the X-ray flux: $L_X \lesssim 2 \times 10^{35} \erg \s^{-1}$.
\cite{Beasoretal2026} point out similarities to erupting stars thought to be stellar mergers.  The luminosity does not decrease much at long wavelengths, leading them to conclude that the longer-wavelength observations are not consistent with an overall decrease in the system's bolometric luminosity.
\cite{Beasoretal2026} find that the obscuring dust is non-spherically distributed, and from that, conclude that the inferred bolometric luminosity is only a lower limit, because a large fraction of the central source’s radiation may escape without being reprocessed by dust. This is the claim of the type II ILOT scenario \citep{KashiSoker2017, SokerTypeII2021}. 
Overall, \cite{Beasoretal2026} present strong arguments against the failed-supernova scenario for M31-2014-DS1 from comparing observations to the failed-supernova scenario predictions. These arguments add to the earlier arguments against this scenario that \cite{Soker2024UnivReview} brought (listed here in Section \ref{sec:intro}).
\cite{DeKetal2026} estimate the ejected gas to be of $\approx 0.1 M_\odot$ and expanding at a velocity of $\simeq 100 \km \s^{-1}$. This gives a kinetic energy of $\approx 10^{46} \erg$. This is comparable to the radiated energy by fallback accretion in their model, which is $E_{\rm fb} \approx 10 \yr \times 10^4 L_\odot$. 
In this study, I present additional problems with the failed-supernova scenario discussed by \cite{DeKetal2026}. 

If the pre-collapse progenitor of M31-2014-DS1 would rotate rapidly, then the accreted inner envelope would form an accretion disk around the newly born black hole. This accretion disk would launch energetic jets (or disk winds) that would drive an energetic CCSN (e.g., \citealt{Gilkisetal2016}). I consider a significant accretion disk to be formed at a radius somewhat larger than the innermost stable circular orbit, say at $r \simeq 5 R_{\rm BH}$, where I take here the black hole radius $R_{\rm BH}$ to be the Schwarzschild radius. To prevent the formation of such an accretion disk that would launch energetic jets, the specific angular momentum of the accreted gas should be $j_{\rm acc} \lesssim 7 \times 10^{16} \cm^2 \s^{-1}$, where I take the black hole mass as $M_{\rm BH} = 5 M_\odot$ as \cite{DeKetal2026} do. For material accreting from the inner envelope at radius $r_{\rm e,acc}$, the rotation velocity at this radius and the rotational period should be $v_{\rm rot} \lesssim 1 \km \s^{-1} (r_{\rm e,acc}/10 R_\odot)^{-1}$ and  $P_{\rm rot} \gtrsim 1.4 (r_{\rm e,acc}/10 R_\odot)^{2} \yr$, respectively. This is not a strong constraint on a yellow supergiant with a radius of $R\simeq 500 R_\odot$, even if $r_{\rm e,acc}$ is an order of magnitude larger. 
 
\begin{figure}
\begin{center}
\includegraphics[trim=3.0cm 9.1cm 0.0cm 9.5cm ,clip, scale=0.56]{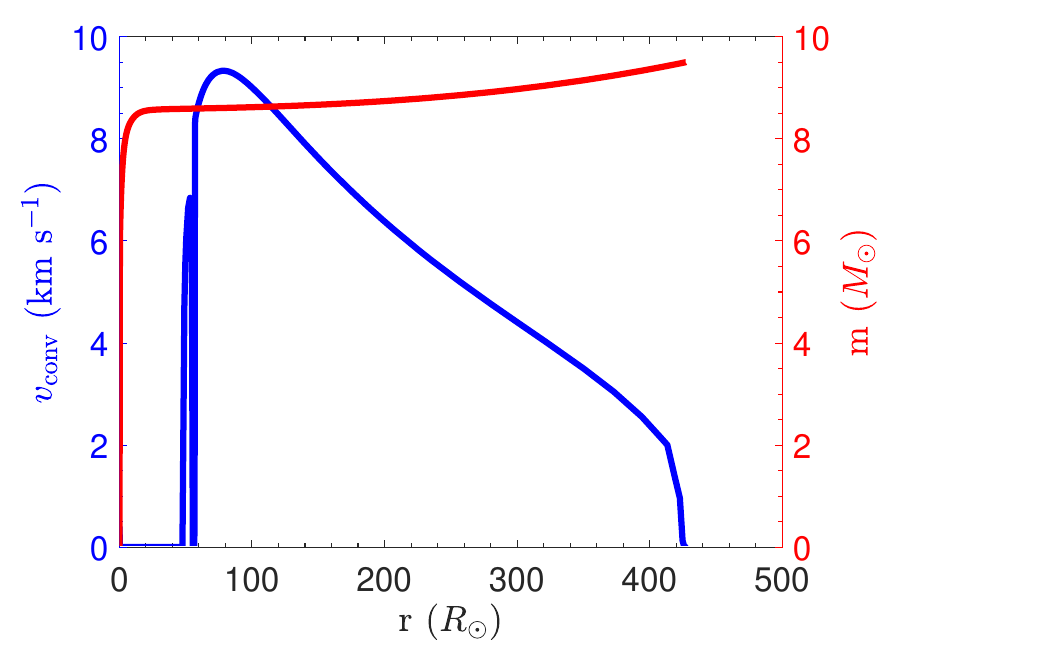}
\caption{A stellar model of a post-red-supergiant star, a yellow supergiant, which was stripped of most of its hydrogen-rich envelope (from \cite{CohenBearSoker2026}). The initial stellar mass was $M_{\rm ZAMS}=20 M_\odot$. At the time shown, the total mass is $M_\ast= 9.5 M_\odot$, the total hydrogen mass is $M_{\rm H}=1.6 M_\odot$, the stellar luminosity is $L_\ast=7.9 \times 10^4 L_\odot$, the effective temperature is $T_{\rm eff}=4680 \K$, and its radius is  $R_\ast=427 R_\odot$. The blue line is the convective velocity $v_{\rm conv}(r)$, while the red line is the mass $m(r)$.       
}
\label{fig:Star}
\end{center}
\end{figure}

Consider then that the specific angular momentum due to pre-collapse rotation in the entire envelope is not sufficient to form an accretion disk around the black hole. In these cases, the stochastic specific angular momentum of the convective cells might lead to the formation of intermittent accretion disks (e.g., \citealt{PapishSoker2011, GilkisSoker2014, ShishkinSoker2021, AntoniQuataert2022, AntoniQuataert2023, WangShishkinSoker2024, WangShishkinSoker2025}), that might launch jets (e.g., \citealt{PapishSoker2014Planar, ShishkinSoker2022}). 
In Figure \ref{fig:Star}, I present the convective velocity and mass as a function of radius for an evolved yellow supergiant (post-red supergiant) star with a zero-age main-sequence mass of $M_{\rm ZAMS}=20 M_\odot$ that was striped from most of its hydrogen-rich envelope (simulated by \citealt{CohenBearSoker2026}). The specific angular momentum parameter that the JJEM uses (e.g., \citealt{WangShishkinSoker2025}), $j_{\rm conv} \equiv v_{\rm conv} r$, characterizes the specific angular momentum fluctuations. In the outer envelope, its value is $j_{\rm conv} \simeq 7 \times 10^{18} \cm^2 \s^{-1}$. This implies that the typical initial radius of the disk (or ring) of the fallback gas around the black hole due to the convective angular momentum fluctuations is 
\begin{equation}
R_{\rm {\rm d,0}} \simeq  1.06   
\left( \frac{ j_{\rm conv}}{7 \times 10^{18} \cm^2 \s^{-1}} \right)^2 
\left( \frac{ M_{\rm BH}}{{5 M_\odot}} \right)^{-1} R_\odot .    
\label{eq:Rdisk}    
\end{equation}

\cite{DeKetal2026} claim that $\gtrsim 99\%$ of the initially bound material will be expelled due to the random angular momentum barrier near the BH horizon. It seems that this is unavoidable from momentum and energy considerations. 
In the scenario described by \cite{DeKetal2026}, the fallback accretion rate decreases as $\propto t^{-2}$. Consider a time longer than the collapse time of the progenitor, which is $0.28 \yr$ for a star of $5 M_\odot$ and a radius of $500 R_\odot$. The amount of mass accreted from $t=0.5 \yr$ to $t=1 \yr$ equals the amount of mass that will be accreted after one year. Namely, in this half a year, the amount of mass accreted is the amount of mass still bound to the black hole. The material that falls-back after a year reaches a maximum distance from the $M_{\rm BH} = 5 M_\odot$ black hole in their scenario of $R_{\rm max} > 800 R_\odot$, implying a specific binding energy per unit mass (defined positively) of 
\begin{equation}
e_{\rm {b,1}} <  1.2 \times 10^{13} \erg \g \simeq 
- \frac{1}{2} \left( 50 \km \s^{-1} \right)^2 .    
\label{eq:ebound1}    
\end{equation}
As stated, the accretion onto the black hole is via an accretion disk.  
Consider that the accretion disk launches jets (or a disk-wind) from a typical radius of $5 R_{\rm BH}$, somewhat larger than the innermost stable circular orbit, and at the orbital velocity at that radius. Namely, $v_{\rm jet} \simeq 10^5 \km \s^{-1}$. 
This velocity is an underestimate because the jets may be relativistic. 
For example, relativistic jets are common in X-ray binaries (e.g., \citealt{McClintockRemillard2006}). For the present study, non-relativistic jets or disk-winds are sufficient. 
Consider that the accretion disk launches a fraction $f_{\rm j} \equiv \dot M_{\rm jets} / \dot M_{\rm acc}$ of the accreted mass, with typical values of $f_{\rm j} \simeq 0.001-0.1$.   
The ratio of the jets' energy in the half-year accretion to the binding energy of the remaining bound material is therefore
\begin{equation}
\frac{E_{\rm jets}}{E_{\rm b,1}} > 4000
\left( \frac{f_{\rm j}}{10^{-3}} \right) 
\left( \frac {v_{\rm jet}}{10^5 \km \s^{-1}} \right)^2 . 
\label{eq:Eratio}    
\end{equation}
However, radiative cooling is fast in the interaction zone of the jet and the bound gas (see Section \ref{sec:Xray}). Therefore, we need to consider momentum conservation. In this example of equal masses of the accreted gas in half a year and the still-bound gas, the post-collision velocity is    
\begin{equation}
v_{\rm b,1} \approx 50 \km \s^{-1} 
\left( \frac{f_{\rm j}}{10^{-3}} \right) 
\left( \frac {v_{\rm jet}}{10^5 \km \s^{-1}} \right) . 
\label{eq:Pvel}    
\end{equation}
This is about the escape velocity from $r= 800 R_\odot$, the radius from which material falls after a year.  

The conclusion is that, indeed, for typical parameters of jet-launching by disks, in the case of a fallback material with stochastic angular momentum (hence the jets are launched in different directions), the initial accretion is expected to launch jets that expel most of the bound mass, preventing efficient late accretion. This is compatible with the claim of \cite{DeKetal2026}.  
However, there are two points to consider here. (a) For a more typical value of $f_{\rm j} \simeq 0.01-0.1$, and for an efficiency somewhat larger than the purely momentum conserving, i.e., efficiency between energy conserving in equation (\ref{eq:Eratio}) and momentum conserving in equation (\ref{eq:Pvel}), I expect that the fallback material in the first several months in their scenario will completely shut down the backflow accretion. (b) As I argue in Section \ref{sec:Xray}, this inefficient accretion of initially bound material yields a very bright source that is incompatible with observations. 

\section{Luminosity constraints}
\label{sec:Xray}

In the failed-supernova scenario that \cite{DeKetal2026} present, there is a gas of $\simeq 0.1 M_\odot$ in a shell between $r\simeq 40 \AU$ and $r \simeq 80 \AU$. 
The average hydrogen number density of this gas is $n_{\rm H} \simeq 10^{10} \cm^{-3}$. 
The bound mass is lower, $M_{\rm b} \simeq 0.05 M_\odot$, but it is at smaller radii, so its density is larger (and its radiative cooling time shorter). 
The interaction of this gas with an outflowing jet or wind will heat it to X-ray-emitting temperatures. Using a cooling function at the relevant temperature (e.g., \citealt{Gaetzetal1988}), $\Lambda \gtrsim 3 \times 10^{-23} \erg \cm^3 \s^{-1}$, the radiative cooling time at constant pressure of this gas is $\tau _{\rm cool} (T) \simeq (5/2) (nkT)/(n_e n_{\rm H} \Lambda)$, where $n_e$ is the electron number density and $n$ total number density. Substituting typical values for the studied case yields 
\begin{equation}
\tau _{\rm cool}  \lesssim 
0.007 \left( \frac{T}{10^8 \K} \right) 
\left( \frac{N_{\rm H}}{10^{10} \cm^{-3}} \right)^{-1} 
\yr . 
\label{eq:TauCool}  
\end{equation}

The conclusion is that most of the interaction energy released in the interaction of the outflowing gas from the black hole vicinity with the bound gas is radiated, rather than converted to kinetic energy. This energy can be very large if the jets (or disk wind) originate from several tens of the Schwarzschild radius or less. \cite{DeKetal2026} mentions the centrifugal barrier that makes the accretion highly inefficient. In that case, most of the backflowing material is ejected from the initial disk (or ring) radius given by equation (\ref{eq:Rdisk}). As usual in outflows from astrophysical objects, the velocity of this gas will be the escape velocity from this radius. I take the terminal velocity (at large distances) of the outflow to be the Keplerian velocity of the disk at its radius, and use the same assumptions that lead to equation (\ref{eq:Rdisk}), yielding    
\begin{equation}
\begin{split}
v_{\rm d,wind} & \simeq \frac{ G M_{\rm BH}}{ j_{\rm conv} }= 
948 \left( \frac{ M_{\rm BH}}{{5 M_\odot}} \right)
\\ & \times 
\left( \frac{ j_{\rm conv}}{7 \times 10^{18} \cm^2 \s^{-1}} \right)^{-1} 
\km \s^{-1}. 
\label{eq:vdiskwind} 
\end{split}
\end{equation}

Taking the bound mass as in \cite{DeKetal2026}, $M_{\rm b} \simeq 0.05 M_\odot$, and taking most of it to be ejected from the disk radius (the centrifugal barrier) within the 10 years since the fading started, the average power of the outflow in these 10 years is 
\begin{equation}
\begin{split}
\bar {\dot E}_{\rm d,wind} & \simeq \frac{1}{2} \frac{M_{\rm b} v^2_{\rm d,wind}}{10 \yr}  
= 3.7 \times 10^5 
\\ & \times   
\left( \frac{M_{\rm b}}{0.05 M_\odot} \right) 
\left( \frac{v_{\rm d,wind}}{ 950 \km \s^{-1}} \right)^2 
L_\odot. 
\label{eq:Ediskwind} 
\end{split}
\end{equation}
This luminosity is a lower limit because there is an energy dissipation within the accretion ring/disk at the radius of $R_{\rm d,0} \simeq 1 R_\odot$ before the gas is ejected, some mass will end in a disk at a smaller radius as the angular momentum fluctuations have lower values of angular momentum, and some mass accretes onto the black hole via an accretion disk very close to the black hole, which launches jets at much higher velocities.  This outflowing gas collides with still-bound gas and the gas that was ejected and has slower velocities of $\approx 100 \km \s^{-1}$, converting most of the kinetic energy to radiation. 
Overall, the luminosity from outflow interaction with the outer material will be an order of magnitude higher than the average observed luminosity and likely substantially higher.

\section{Summary} 
\label{sec:Summary}

I examined the failed-supernova scenario proposed by \cite{DeKetal2026}, based on \cite{Deetal2024}, to explain the fading of M31-2014-DS1. In Section \ref{sec:Jets}, I examined the expected outflow from the accretion disk that the fallback gas forms around the black hole. I conclude that the jets or disk wind launched by the fallback material in the first year or so are likely to completely shut down fallback accretion. The source of angular momentum is the stochastic pre-collapse convective motion in the envelope. This will lead to jittering jets that expel material in many directions, efficiently ejecting the outer-bound material.  
\cite{DeKetal2026} requires that most of the bound gas is ejected back because of the angular momentum barrier. I find that the fallback gas, with the typical angular momentum fluctuations as inferred from the stellar model I presented in Figure \ref{fig:Star}, will form an accretion disk at $R_{\rm {\rm d,0}} \simeq 1 R_\odot$ (equation \ref{eq:Rdisk}). Assuming that the gas leaves this radius at a terminal velocity of about its Keplerian rotation velocity, the average power of the outflow is given by equation (\ref{eq:Ediskwind}). Due to the fast radiative cooling time of the interacting outflows (equation \ref{eq:TauCool}), most of this energy is radiated away. This will give a luminosity about an order of magnitude larger than the observed luminosity.  
 
Overall, I cannot completely rule out the failed-supernova scenario proposed by \cite{DeKetal2026}. I only argue that for it to work, it requires fine-tuning parameters that are not typical for astrophysical objects. An example of fine-tuning is that the jets (or disk wind) of the fallback gas that forms accretion disks will expel $>99 \%$ of the bound material, but will not completely eject it for 10 years or more. An example of an atypical value is that the fraction of the accretion disk mass that the black hole launches should be very low. For example, if only $0.1 \%$ of the bound mass is accreted onto the black hole (a fine-tuned value), namely, $ 5 \times 10^{-5} M_\odot$, and it launches jets that carry typical values of $\simeq 1 \%$ of the rest energy of the accreted gas, then the total energy the jets carry is $\approx 10^{48} \erg$, much larger than the total radiated energy in the decade of the fading event ($< 10^{47} \erg$). To prevent this large luminosity, the jets (or disk wind) from the disk around the black hole should carry a very small fraction of energy. \cite{DeKetal2026} noted some of these atypical values.      

The specific problems I identify in the failed-supernova scenario for M31-2014-DS1 in this study add to the general problems I raised before \citep{Soker2024UnivReview} and listed in Section \ref{sec:intro}, as well as to the problems \cite{Beasoretal2026} discuss in relation to their new observations. These problems and challenges lead me to consider the failed supernova scenario unlikely for M31-2014-DS1. 
I conclude that there is no justification for the claim of \cite{DeKetal2026} that their analysis of M31-2014-DS1 provides the first cohesive insights into black hole formation via low-energy explosions and long-term fallback. \cite{Beasoretal2026} approach of considering both the failed-supernova and the type II ILOT scenarios is the correct one. 

More generally, the community should discuss both CCSN explosion mechanisms, the JJEM that predicts no failed supernovae, and the neutrino-driven mechanism that predicts failed supernovae. In the neutrino-driven mechanism, some models do not explode, leading to black hole formation without a supernova (failed supernova). In the JJEM, an inefficient jet-feedback mechanism in cases of rapid pre-collapse core rotation with large binding energy (massive stars) leads to black hole formation (e.g., \citealt{Soker2023gap}) but also to a very energetic explosion (e.g., \citealt{Gilkisetal2016}). 

\section*{Acknowledgements}

I thank an anonymous referee for their supportive comments, Ealeal Bear for the figure, and the Charles Wolfson Academic Chair at the Technion for the support.


%
\bibliography{reference}{}
\bibliographystyle{aasjournal}
  


\end{document}